# DISTRIBUTED CAN-BUS BASED BEAM DIAGNOSTIC SYSTEM FOR PULSE RACE-TRACK MICROTRON

F. Nedeoglo, O. Novojilov, S. Dudnikov, Department of Physics, Moscow State University, 119899, Moscow, Russia

A. Chepurnov, I. Gribov, V. Shvedunov, Institute for Nuclear Physics, Moscow State University, 119899, Moscow, Russia

## Abstract

Very compact 70 MeV pulse race-track microtron is under construction now. To acquire outputs of beam-current transformers on every orbit and pulses of high voltage and RF field a distributed multi-channel beam diagnostic system was developed. Each acquisition controller consists of four fast differential amplifiers and one DSP-based micro-controller with on-chip ADC and CAN-bus controller. Each amplifier is coupled with beam-current transformer and has bandwidth of up to 150MHz and gain of up to 10. One of four channels is acquired during a measurement cycle. Another channel could be selected between two following pulses. All the controllers are connected via optically coupled CAN-bus with a host diskless PC running under Linux with the RTLinux extension. Dedicated software of the system consists of low level acquisition software for DSP, network software for controllers and host PC, application software for PC to present date for operator and control system. Standard CAN application layers were considered but refused because of the closed character of the whole system and centralised synchronisation of the whole system.

## 1 INTRODUCTION

First successful runs of the very compact pulse 70 MeV race-track microtron (RTM) have been provided and final tuning is carried out now. [1]. Parameters of the RTM are listed in the Table 1.

Table 1.

| Injection energy | 50 keV |
|---|---|
| Energy gain/orbit | 5 MeV |
| Output energy | 10-70 MeV |
| Number of orbits | 14 |
| Output current at 70 MeV | 40 mA |
| Pulse length | ~6-10 µs |
| Pulse repetition rate | 150 Hz |
| Dimensions | 2.2x1.8x0.9 mm |
| Weight | 3200 kg |

Because of the limited place between orbits, the original small size pulse beam current monitor (BCM) has been designed. The BCM is a passive wide-band current transformer with sensitivity up to 4,9 V/A and double-ended 50 Ohm-coupled output.

To measure amplitudes of the beam current in each orbit together with the amplitude of RF-field and high voltage pulse, a multi-channel distributed data acquisition beam diagnostic system has been created.

## 2 BEAM DIAGNOSTIC SYSTEM AS A PART OF CONTROL SYSTEM

The diagnostic system provides data necessary for control algorithms and human-machine interface (HMI) which are implemented in control system (CS) of the accelerator, therefore the system has been designed in such a way to be easily integrated with CS.

CS has a traditional three level structure [2]. X86-compatible computers are used. Front-end level consists of diskless PC with data acquisition boards. Middle level consists of diskless PC running under Linux together with real-time extension of the Linux - RTLinux. Linux is used to implement static and soft real-time algorithms whereas RTLinux is used to run hard real-time algorithms. HMI and the data bases are implemented in the third level. Ethernet over fibre optic is used to connect PCs in the accelerator hall with servers and HMI computers in the control room.

Beam diagnostic looks from top level of CS like one more dedicated acquisition subsystem but has different implementation architecture of front-end level.

## 3 STRUCTURE OF THE BEAM DIAGNOSTIC SYSTEM

The following technologies developed during the last few years have been used for the system: application of diskless PC running under Linux with real-time extension - RTLinux [3,4]; application of distributed stand-alone DSP- based smart controllers [5]; application of CAN-bus for accelerator control [6].

The output signal of BCM is measured by a stand-alone intelligent controller. Every controller has four inputs for the BCM. One of the four amplified signals could be digitised in a single acquisition cycle. The digitising process is synchronised by a dedicated pulse generated by the general synchronisation system of the RTM. CAN-bus is used to connect controllers with the diskless x86-compatible host computer running under Linux together with real-time extension RT-Linux (Figure 1.). BOOTP protocol is used to download the operating system to the host computer via Ethernet after switching power on. Host computer is equipped with an in-house designed CAN-bus adapter [5].

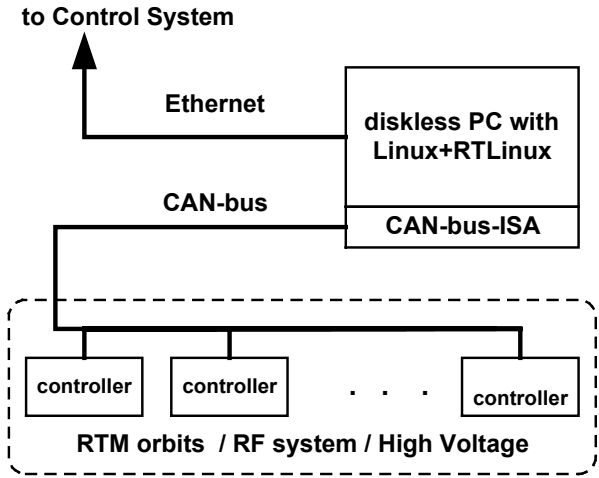

Figure 1. Structure of beam diagnostic system.

CAN-bus is a very popular fieldbus for accelerator control. Proceedings of ICALEPCS, PCAPAC and other conferences shows growing the popularity of CAN-bus for accelerator control with every next year. Maximum speed of CAN-bus is 1 Mbit/sec. But it is enough when the beam diagnostic system is used to measure values necessary for relatively slow static algorithms and HMI implemented in the high level of CS. The pulsed nature of the data allows transfer of the data in time gaps between two following pulses.

## 4 ACQUISITION CONTROLLER

The acquisition controller consists of analogue and digital parts (Figure 2). Four independent fast instrumental amplifiers (IA) are implemented in the analogue part. Every IA has unit gain bandwidth up to 200 MHz. Each of the IA could be separately enabled or disabled by the controlling DSP. All outputs of the IA are connected together to the inputs of two additional buffer amplifiers. They are used to couple the output of the IA with the ADC input and test analogue output simultaneously. The test analogue output allows us to use a digitising oscilloscope to measure and store the shape of the pulses in each orbit of the RTM.

The digital part consists of a digital signal processor (DSP) TMS320F241, an optically decoupled CAN-bus interface, an optically decoupled synchronisation input, a synchronisation and control schematic based on CPLD and an RS-232 interface. The DSP has an on-chip CAN-bus controller, fast ADC and other useful peripherals. The fast on chip ADC has 10-bits resolution and an 800 ns minimum conversion time.

Interrupt service mode of DSP operation allows us to utilise the high performance of DSP and ADC module.

The synchronisation pulse coupled with the beam pulse starts data acquisition process. One of the four channels is measured during one measuring cycle. The host computer sets the number of the channel to measure after the next synchronisation pulse comes. In addition, the host computer checks the state of the controller, defines the number of continuous measurements and initiates transmission of the results from measuring controller to the host via CAN-bus.

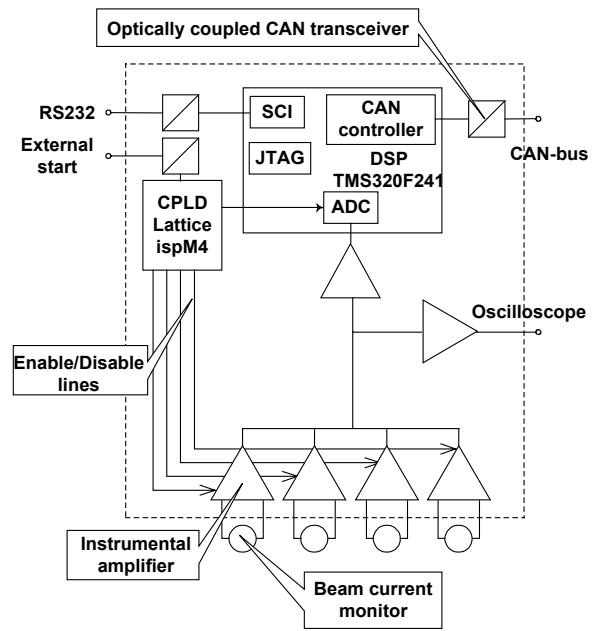

Figure 2. Structure of acquisition controller.

## 5 BEAM DIAGNOSTIC SOFTWARE

Standard CAN application layers such as CANopen and DeviceNet were considered as candidates for CAN application layers for the beam diagnostic system. Because of the following reasons the dedicated high level CANdiag protocol was created:
- the diagnostic system is closed to future extension, so a custom protocol is acceptable;
- DSP has limited size of on chip Flash-memory that is too small for standard protocols. Application of external memory is not reasonable;
- centralised synchronisation of the system and asymmetric flows of data makes application of standard protocols inconvenient.

The CANdiag protocol is based on a master-slave model of interaction. The master portion of protocol is implemented in host computer whereas all controllers are slaves.

Only the 11-bit CAN-identifier is used. Figure 3 represents usage of CAN-identifier bits.

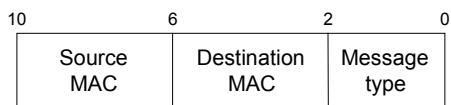

Fig. 3. Usage of CAN-identifier field in CANdiag protocol.

Each controller connected to the CAN-bus has its unique Media Access Control (MAC) identifier (ID), which identifies the device in the network and is used during the procedure of access to the bus. When the device sends a message to the bus, the first four bits of the CAN-identifier contain the MAC ID of sender (Source MAC ID). The next four bits contain the MAC ID of the device which expected to be receiver of this message (Destination MAC ID). The last three bits are used to identify the type of the message which define the semantic meaning of the message and format of data fields.

The CANdiag protocol supports simultaneous operation up to 15 devices in CAN-bus with addresses in the range between 0x00 and 0x0e. MAC ID 0x0f is used for broadcast messages to implement duplication MAC ID checking. Each device on CAN-bus working under CANdiag protocol starts its activity on the bus after switching power on with the duplication MAC ID checking procedure. The CAN-bus node sends broadcast message with Source MAC ID equal to Destination MAC ID and equal to 0x0f. Data field contains MAC ID of the node trying to connect to the bus. All nodes of the bus which are active in this moment receive this broadcast message and compare the MAC ID from data field with its own MAC ID. If the received MAC ID is equal to its own MAC ID, this node sends a broadcast reply which means that the requested MAC ID is occupied already and it means that the attempt to connect to the bus failed.

The CANdiag protocol supports the following types of messages defined by "Message type" field:
- configuration messages – are used to select dedicate measurement channel in slave device and reset controller remotely,
- status messages are used to check state of the controller;
- input/output messages are used to transfer stored digitised data.

Application software of the system was developed in ANSI C and consists of the low level software of the slave controller running on the DSP and the high level software of master.

Software for slave part of the CANdiag protocol was completely created, tested and debugged under Linux in emulation mode taking in to consideration features of the C –compiler for the DSP platform. Then pieces of code were ported very easily and quickly to the DSP.

Software of the master consists of a loadable module for RTLinux 3.0 and application software running under Linux on the same host computer. Application modules allow scanning of the CAN-bus to check state of all slave controllers, to provide cyclic polling of the controllers and so on.

A dedicated API is used between the host computer and the general CS of RTM to allow access to the beam diagnostic system from CS. The control program that is a part of the CS software uses two real-time FIFOs to communicate with the master's software. One FIFO is used to transmit commands from the control program to the master's program. The second FIFO is used to transfer results back from the master to the control program.

## 4 CONCLUSIONS

To simplify unification of the beam diagnostic system with the control system during start up and future operation the same architectural decisions should be used. Single platform of software development consisting of GNU C under Linux together with RTLinux was used. The platform was used to develop software for the CS, for the beam diagnostic system, as well as for high level, for embedded applications, and for real-time as well as for non real-time components. This approach is very convenient and could be recommended to develop control and beam diagnostic systems. One more application of CAN-bus for beam diagnostic systems is described. A disk-less PC running under Linux could be recommended as reliable and inexpensive solution for middle level of control systems.

## REFERENCES

[1] V.I. Shvedunov, et. al., "70 MeV Electron racetrack microtron commissioning" Proc. of PAC, Chicago June 18-22, 2001
[2] I.V. Gribov., I.V. et. al., "RaceTrack Microtron Control System" Proc. of PAC, Chicago June 18-22, 2001
[3] A. Chepurnov, F. Nedeoglo, et. al., "Simple CAN-bus adapter for accelerator control running under Linux and RTLinux" CD-ROM Proceedings of PCAPAC'2000
[4] F. Nedeoglo, A. Chepurnov, D. Komissarov, "Linux and RT-Linux for accelerator control - pros and cons, application and positive experience" Proc. Of ICALEPCS'99, Trieste, Italy, ISBN: 88-87992-00-2, pp. 520-522.
[5] Chepurnov A.S., Dorokhin A.A., et. al., "Control System for Accelerator with distributed Intelligence Based on a "Family of Smart Devices", Proc. of the Vth Europ. Particle Accelerator Conference Sitges (Barcelona), Institute of Physics Publish. Bristol and Philadelfia, 1996, p.1794-1796
[6] A. Chepurnov, A. Alimov, et. al., "Control System for New Compact Electron Linac" Proc. Of ICALEPCS'99, Trieste, Italy, ISBN: 88-87992-00-2, pp.84-86.